\documentclass[11pt]{interact}

\usepackage{epstopdf}
\usepackage[caption=false]{subfig}

\usepackage[numbers,sort&compress,merge]{natbib}
\bibpunct[, ]{[}{]}{,}{n}{,}{,}

\usepackage{color}

\theoremstyle{plain}

\theoremstyle{definition}

\theoremstyle{remark}

\graphicspath{{figs/}}

\begin{document}

\articletype{to appear in \it{Molecular Physics}}

\title{Structure and equation-of-state of a
  disordered system of shape anisotropic patchy
  particles}

\author{ \name{Susanne Wagner and Gerhard
    Kahl \thanks{contact: G. Kahl. (gerhard.kahl@tuwien.ac.at)}}
\affil{Institut f\"ur Theoretische Physik and
  Center for Computational Materials Science (CMS), TU~Wien, Wiedner
  Hauptstra{\ss}e 8-10, A-1040 Wien, Austria}}

\maketitle

\begin{abstract}
We present a new two dimensional model for elliptic (i.e., {\it shape
  anisotropic}) patchy colloids, where the impenetrable core of the
particles is decorated on its (co-)vertices by Kern-Frenkel type of
patches. Using (i) well-documented criteria for the overlap of the
undecorated ellipses and (ii) proposing new criteria that give
evidence if the patchy regions of two particles interact we perform
extensive Monte Carlo simulations in the NPT ensemble. Considering
elliptic particles of aspect ratios $\kappa =$ 2, 4, and 6 with two
patches located at the co-vertices and choosing representative values
for the temperature we study on a semi-quantitative level the emerging
disordered phases in terms of snapshots and radial distribution
functions and present results for the equation-of-state.
\end{abstract}


\section{Introduction}
\label{sec:introduction}

Patchy particles, i.e. colloids that are decorated on their otherwise
spherical (or circular) surface by regions of repulsion or attraction
have been subject of a steadily increasing number of experimental and
theoretical investigations during the past one to two decades. This
remarkable interest in these particles is to a considerable amount due
to the fact that patchy particles can meanwhile be synthesized with
essentially arbitrary patch decorations in terms of number, positions,
and spatial extent of the patches. They have therefore become very
versatile units in bottom-up self-assembly strategies of soft matter
based functional materials. Reviews over and special issues
dedicated to patchy particles covering both experiment and theory
provide an excellent overview over this rapidly expanding field
\cite{Bianchi:2011,Bianchi:2017,COCIS:2017,Bianchi:2017a}.

In this contribution we put forward a novel model for {\it shape
  anisotropic} patchy particles. With such a concept we give due
credit to recent successes in synthesizing particles with shapes of
ever increasing complexity. For an overview see, for instance,
\cite{Boles:2016} and references therein.

As it is difficult to grasp both shape anisotropy {\it and} patchiness
at the same time using simulations or theoretical approaches we put
forward in this contribution a rather simple model which nevertheless
fully captures the two aforementioned features: we consider
impenetrable (two dimensional) elliptic particles of aspect ratio
$\kappa$ which can be decorated on their (co-)vertices with patches;
these interaction sites are realized by the widely used Kern-Frenkel
model \cite{Kern:2003}, namely patches located on a circle (or
sphere), specified by an opening angle and a finite interaction range.
Within the patch region the potential is assumed to be constant (i.e.,
a square-well type of interaction).  Depending on the sign of the
interaction parameter, the patches can be attractive or repulsive.

For using this model in Monte Carlo (MC) simulations we had to find
criteria which provide evidence if two of the anisotropic patchy
particles interact: (i) for the undecorated, impenetrable elliptic
bodies we use the criterion put forward by Vieillard-Baron
\cite{Vieillard-Baron:1972,Vieillard-Baron:1970}, indicating if two
ellipses (with given orientations of their directors, ${\bf u}_1$ and
${\bf u}_2$, and separated by a given center-to-center vector ${\bf
  r}$) overlap, touch each other, or do not overlap; (ii) in an effort
   to identify possible patch-patch interactions
  we map via the osculating circles of the (co-)vertices our model on
  the original Kern-Frenkel model for circular (or spherical) patchy
  particles \cite{Kern:2003}.

MC simulations are carried out in the NPT ensemble, allowing for
variations in the area and in the shape of the simulation box;
ensembles of 1260 particles have been considered in all simulation
runs which typically extend over $\sim 1.25~10^5$ cycles. Triggered by
a random number, translational (45\%) or rotational (45\%) moves of
the particles as well as modifications of the box area or shape (10\%)
are carried out alternately.

Similar as in previous studies on undecorated particles
\cite{Cuesta:1990} we have considered three values of $\kappa$, namely
$\kappa = 2, 4$, and 6. Fixing the temperature $T$ to different,
representative values we have systematically varied the pressure
within a range where the disordered phase is stable. Since
  this contribution should rather be considered as a ``proof of
  concept'' to introduce both patchiness {\it and} shape anisotropy in
  one model, we postpone the more demanding (and time-consuming)
  investigations of the emerging {\it ordered} structures and the
  related phase transitions to a forthcoming, more comprehensive
  publication \cite{Wagner:2018}. Configurations have
  been analyzed on a semi-quantitative qualitative level via visual
  inspection and via a (spatial) pair distribution function
  \cite{Hansen:2013}, comparing the emerging phases with previous
studies performed for undecorated, hard ellipses
\cite{Cuesta:1990,Bautista-Carbajal:2014}. Data for the
equation-of-state (i.e., pressure $P$ vs. packing fraction $\eta$)
have been extracted from the simulation data for the selected
$\kappa$-values and for a representative temperatures, including the
high temperature case which corresponds to undecorated, hard ellipses.

The manuscript is arranged as follows: in the subsequent Section we
specify the model and we provide details about the MC simulations. In
Section 3 we present and discuss the results. The manuscript is closed
with concluding remarks and an outlook over future investigations
based on this model.

\section{Model, methods, and analysis}
\label{sec:model_methods_analysis}

\subsection{The model}
\label{subsec:model}

Motivated by the Kern-Frenkel model \cite{Kern:2003}, which was
originally put forward for disk-shaped (or spherical) particles we
have developed the following simple model for a shape anisotropic
patchy particle.

\begin{figure}
\centering
\resizebox*{10cm}{!}{\includegraphics{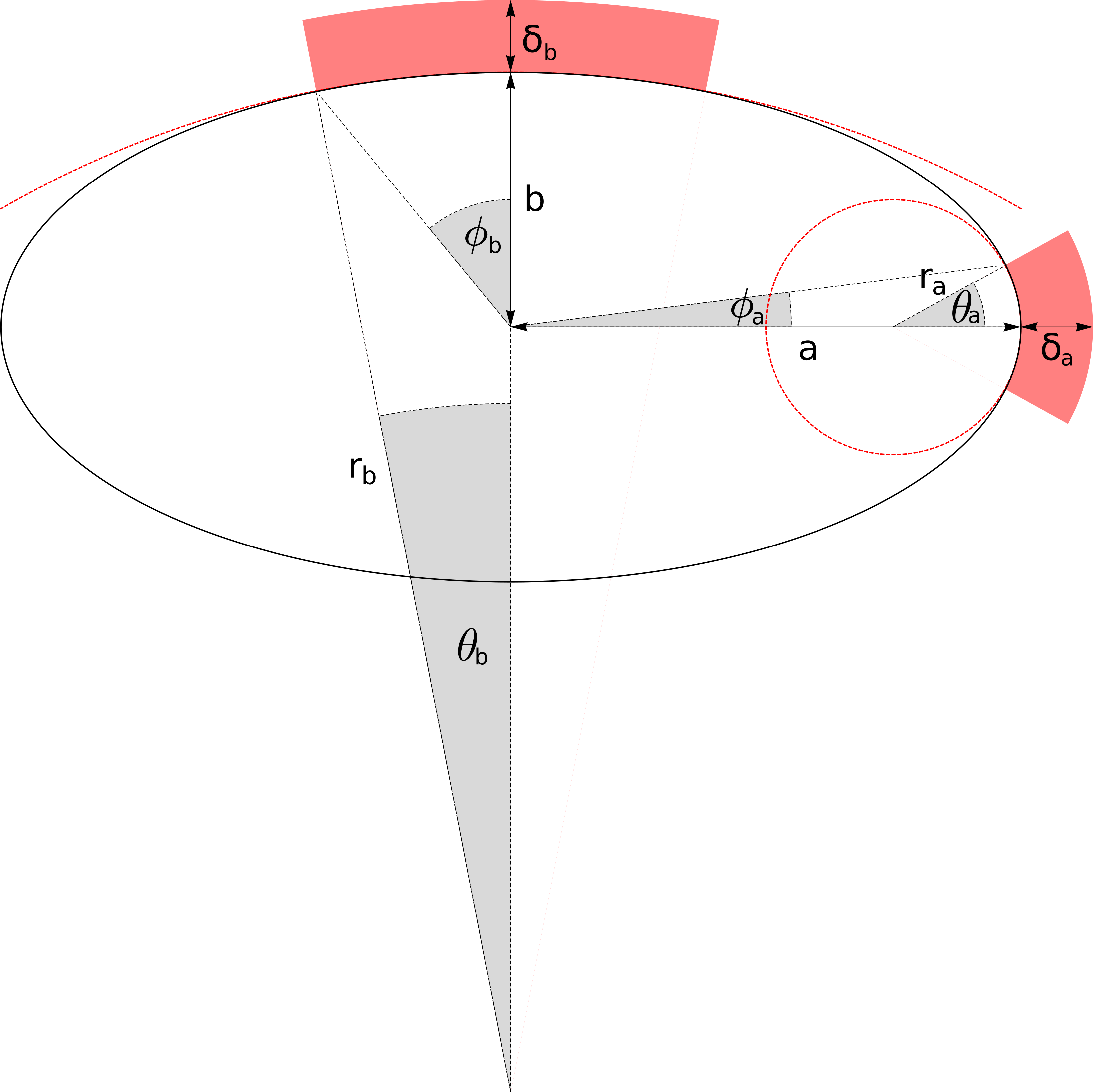}}
\caption{Schematic figure visualizing the geometric definition of the
  patches in our model. For reasons of clarity only two of the up to
  four patches are shown; they can be positioned at the (co-)vertices
  of the elliptic body. The impenetrable, elliptic particle is defined
  via its semi-axes $a$ and $b$; for this particular representation a
  value $\kappa = a/b = 2$ was assumed. The osculating circles of the
  (co-)vertices with radii $r_a$ and $r_b$ are marked by the pink
  lines. The patches (highlighted in pink) are approximated by
  sectorial regions of two concentric circles: the inner one is the
  osculating circle of the respective (co-)vertex while the outer
  circle has a radius which is by $\delta_a$ or $\delta_b$ larger than
  $r_a$ or $r_b$, defining thus the interaction range of the
  patch. Inside the patch region the interaction is assumed to be
  constant. The widths of the patches are measured by the opening
  angles $\theta_a$ and $\theta_b$ (as seen from the centers of the
  osculating circles) or $\phi_a$ and $\phi_b$ (as seen from the
  center of the ellipse).}
\label{fig:single_particle}
\end{figure}

The body of our elliptic particles (with semi-axes $a$ and $b$ and an
aspect ratio $\kappa = a/b$) is impenetrable and can be decorated on
the four (co-)vertices in the following manner by patches (see Figure
\ref{fig:single_particle}; subscripts 'a' and 'b' specify in the
following quantities that are related to the patches located at the
vertices and co-vertices, respectively): the patches are delimited by
a region of two concentric circles, the inner one being the respective
osculating circle. Within a Cartesian coordinate system located in the
center of the ellipse and with the main axis being oriented along the
$x$-axis, the centers of the osculating circles are given by $M_a =
(\pm a\mp r_a,0)$ and $M_b = (0,\pm b\mp r_b)$ with radii $r_a =
a^2/b$ and $r_b = b^2/a$, respectively. The radius of the outer
delimiting circle of a patch is given by $(r_a + \delta_a)$ or $(r_b +
\delta_b)$, setting thus the interaction ranges of the patches. We
assume a square-well type patch interaction with well-depth
$\epsilon_{\rm KF}$, i.e., the interaction strength is set constant
within the patch region as it has been proposed in the original
Kern-Frenkel (KF) model \cite{Kern:2003}.

Approximating the inner limit of the square-well potential by the
osculating circles (instead by the ellipse itself) is, of course,
valid only up to angles $\theta_{a, {\rm max}}$ and $\theta_{b, {\rm
    max}}$ (as seen from the centers of the osculating circles) or
$\phi_{a, {\rm max}}$ and $\phi_{b, {\rm max}}$ (as seen
from the center of the ellipse): imposing via a parameter $\epsilon$
(measured in the unit length scale of the model, i.e., $2
\sqrt{ab}$) a maximum discrepancy between the osculating circle and
the elliptic line (measured along a beam originating in the center of
the osculating circle) defines the limiting angles for the maximum
patch size; of course these limiting angles are
$\kappa$-dependent. Representative values for these quantities are
given in Table \ref{table}, calculated for two different values of
$\epsilon$ (i.e., 0.1 \% and 0.5 \%) and for the three different
$\kappa$-values considered in this study (i.e., $\kappa = 2, 4$, and
6). Further, in this table information about the related maximum patch
extents (for symmetric patches on both (co-)vertices), $p_a$ and $p_b$, 
measured in units of the elliptic circumference, is given.

\begin{table}
\tbl{Values that specify the maximum extent of patches in our model,
  as specified by a criterion put forward in the text: $\theta_{a,
    {\rm max}}$ and $\theta_{b, {\rm max}}$ -- maximum opening angles
  of the patches as seen from the centers of the respective osculating
  circles, $\phi_{a, {\rm max}}$ and $\phi_{b, {\rm max}}$ --
  maximum opening angles of the patches as seen from the center of the
  ellipse, and $p_a$ and $p_b$ -- maximum patch extents as measured in
  units of the elliptic circumference. Representative values for these
  quantities are given for the three aspect ratios $\kappa$ (as they
  are considered in this study) and for two different values of
  $\epsilon$, used as a tolerance in the aforementioned criterion.}
{\begin{tabular}{|l|l||c|c||c|c||c|c|} \hline   
        & & $\theta_{a, {\rm max}}$ & $\theta_{b, {\rm max}}$ & $\phi_{a,{\rm max}}$ & $\phi_{b, {\rm max}}$ & $p_a$ & $p_b$ \\ \hline 
        $\kappa = 2$ & $\epsilon$ = 0.1 \% & $29.30^{\circ}$ & $11.77^{\circ}$ & $7.20^{\circ}$ & $41.69^{\circ}$ & 0.11 & 0.36 \\ 
                     & $\epsilon$ = 0.5 \% & $45.29^{\circ}$ & $17.28^{\circ}$ & $10.86^{\circ}$ & $55.40^{\circ}$ & 0.21 & 0.63 \\ \hline 
        $\kappa = 4$ & $\epsilon$ = 0.1 \% & $36.71^{\circ}$ & $6.03^{\circ}$ & $2.17^{\circ}$ & $61.51^{\circ}$ & 0.04 & 0.40 \\ 
                     & $\epsilon$ = 0.5 \% & $57.60^{\circ}$ & $8.76^{\circ}$ & $3.11^{\circ}$ & $71.55^{\circ}$ & 0.06 & 0.59 \\ \hline 
        $\kappa = 6$ & $\epsilon$ = 0.1 \% & $43.05^{\circ}$ & $4.17^{\circ}$ & $1.09^{\circ}$ & $70.96^{\circ}$ & 0.02 & 0.43 \\ 
                     & $\epsilon$ = 0.5 \% & $67.97^{\circ}$ & $6.04^{\circ}$ & $1.50^{\circ}$ & $78.08^{\circ}$ & 0.03 & 0.62 \\ \hline
\end{tabular}}
\label{table}
\end{table}

As demonstrated in the following, the above specified model is easily
amenable to the usual overlap/interaction acceptance criteria used in
MC simulations:

\begin{itemize}
\item[(i)] following ideas put forward by Vieillard-Baron
  \cite{Vieillard-Baron:1970,Vieillard-Baron:1972} one can formulate
  criteria that are able to provide evidence if two elliptic particles
  (with given orientations and a given center-to-center vector)
  overlap, touch each other, or do not overlap;
\item[(ii)] as a consequence of approximating the patches by circular
  patches on the osculating circles, the overlap criterion used for
  circular Kern-Frenkel type of patches \cite{Kern:2003} can also be
  used in our model: hence two patches interact if the line connecting
  the centers of the related ocsculating circles intersects with the
  patchy regions.
\end{itemize}

\begin{figure}
\centering
\resizebox*{14cm}{!}{\includegraphics{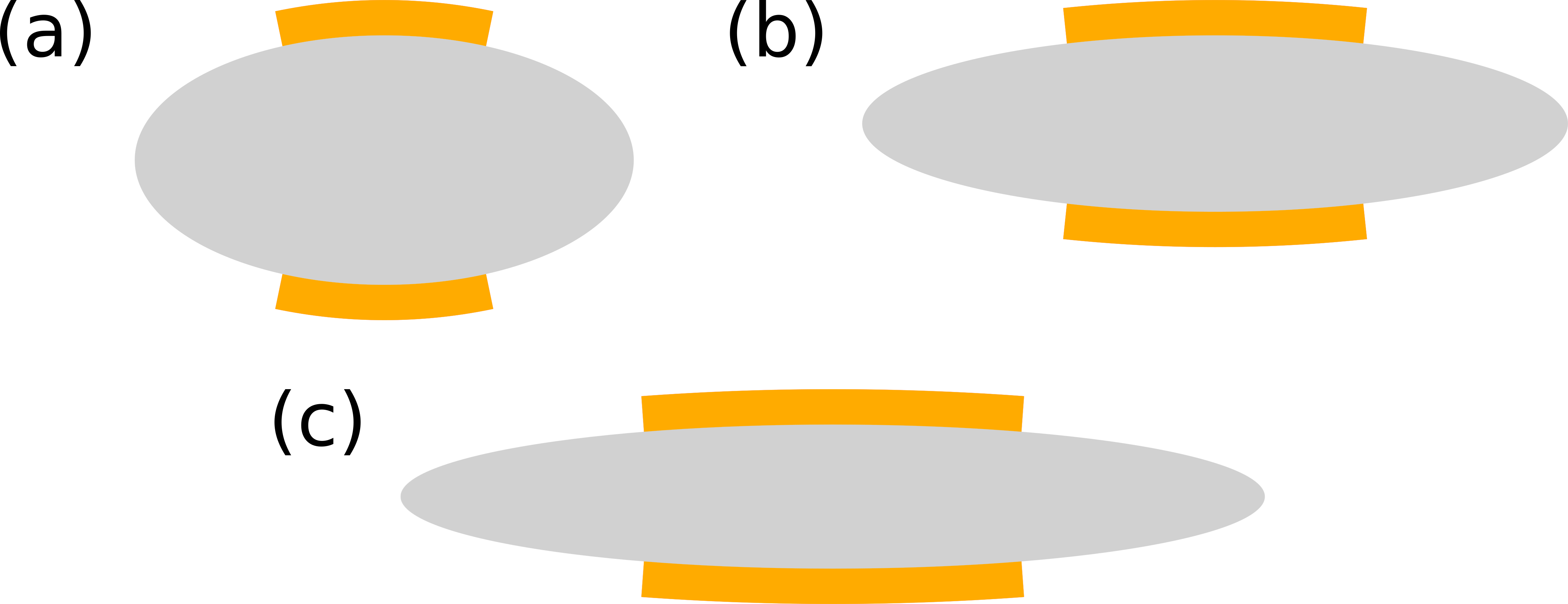}} 
\caption{Schematic representations of elliptic patchy particles as
  they have been used in this contribution; patches are highlighted in
  orange. Particles with aspect ratios $\kappa = 2, 4$ (panels a and
  b), and 6 (panel c) have been considered. In all cases the
  interaction range was chosen to be $\delta_b = 0.1$ (in units of
  $2\sqrt{ab}$); the opening angles $\theta_b=11.77^{\circ},
  6.03^{\circ}$ (panels a and b) and $4.17^{\circ}$ (panel c) are
  chosen such that the criterion for approximating the ellipse by its
  osculating circle within the patch region is fulfilled with an
  accuracy of $\epsilon = 0.1\%$ (see text and Table \ref{table}).}
\label{fig:all_particles}
\end{figure}

\subsection{Monte Carlo simulations}
\label{subsec:MC}

MC simulations were performed in the NPT ensemble, offering thus the
possibility for changes both in the shape as well as in the area of
the simulation box (henceforward termed as ``area moves''). To be more
specific, in an ``area move'' the lengths of the edges of the box are
varied independently; the angle between the edges is not varied (i.e.,
kept at 90$^{\rm o}$) for simulations of the disordered phases, while
for ordered phases also this parameter has to be varied
\cite{Wagner:2018}. The maximum change in the box parameters is
limited in such a way that an overall acceptance rate of 20 - 50 \%
for the ``area moves'' is guaranteed.  Throughout, periodic boundary
conditions were used. Ensembles of $N =$ 1260 particles were
considered and the simulation length extends over 1.25~$10^5$ cycles,
where each cycle consists of $N$ trial moves; triggered via a random
variable each of these moves can either be a translational or a
rotational move of a randomly chosen particle or an ``area move''; the
probabilities of selecting one of these moves are given by 45\% for
the translational and rotational moves and 10\% for the area
moves. Acceptance rules for the three types of moves are based on the
overlap/interaction criteria put forward in Subsection
\ref{subsec:model} and the usual NPT-related selection rules
\cite{Frenkel:2002}. The first 20\% of the cycles were considered as
an equilibration phase and were thus discarded. Each tenth cycle the
configuration was saved resulting in a total of 10$^4$ configurations
available for the analysis.

Simulations were started at low pressure from an initial configuration
which was created as a random particle arrangement (both with respect
to positions and orientations of the particles); the resulting density
was chosen to be close to the one observed in the equation of state in
previous studies on undecorated hard ellipses \cite{Cuesta:1990}.
Subsequent state points were accessed by compressing the preceding
configuration: after an equilibration phase the production run for
this new state point was launched.

The density of the system is measured in terms of the
  packing fraction $\eta$, defined as $\eta = a b \pi N/A$, $A$ being
  the area of the simulation cell. The spatial structure was
  quantified via the conventional pair distribution function, $g(r)$
  \cite{Hansen:2013}. In order to obtain the equation-of-state the
packing fractions $\eta$ have been recorded along the simulation run
(at a given temperature and pressure) and have then been
averaged. The pressure $P$ is given in reduced,
  dimensionless units via $P^\star = \pi a b P /(k_{\rm B} T)$.



\section{Results}
\label{sec:results}

Simulations have been carried out for patchy elliptic
  particles with two patches located on the co-vertices, considering
three different $\kappa$-values, namely $\kappa = 2, 4$, and 6 (see
Fig. \ref{fig:all_particles}). For the range of the patch interaction
we chose $\delta_b = 0.1 \times 2 \sqrt{a b}$. The pressure was varied
over a representative range where we expect (according to previous
results for undecorated ellipses
\cite{Cuesta:1990,Bautista-Carbajal:2014}) the emergence of {\it
  disordered phases}, only. Also the temperature was varied over a
representative range. The patch interaction was assumed to be
attractive and is measured in units of $k_{\rm B}T$, $k_{\rm B}$ being
the Boltzmann constant; the reduced, dimensionless temperature
$T^\star$ is defined as $T^\star = k_{\rm B} T/ \epsilon_{\rm KF}$.

\subsection{Snapshots and positional order}
\label{subsec:structure}

\begin{figure}
\centering
\resizebox*{14cm}{!}{\includegraphics{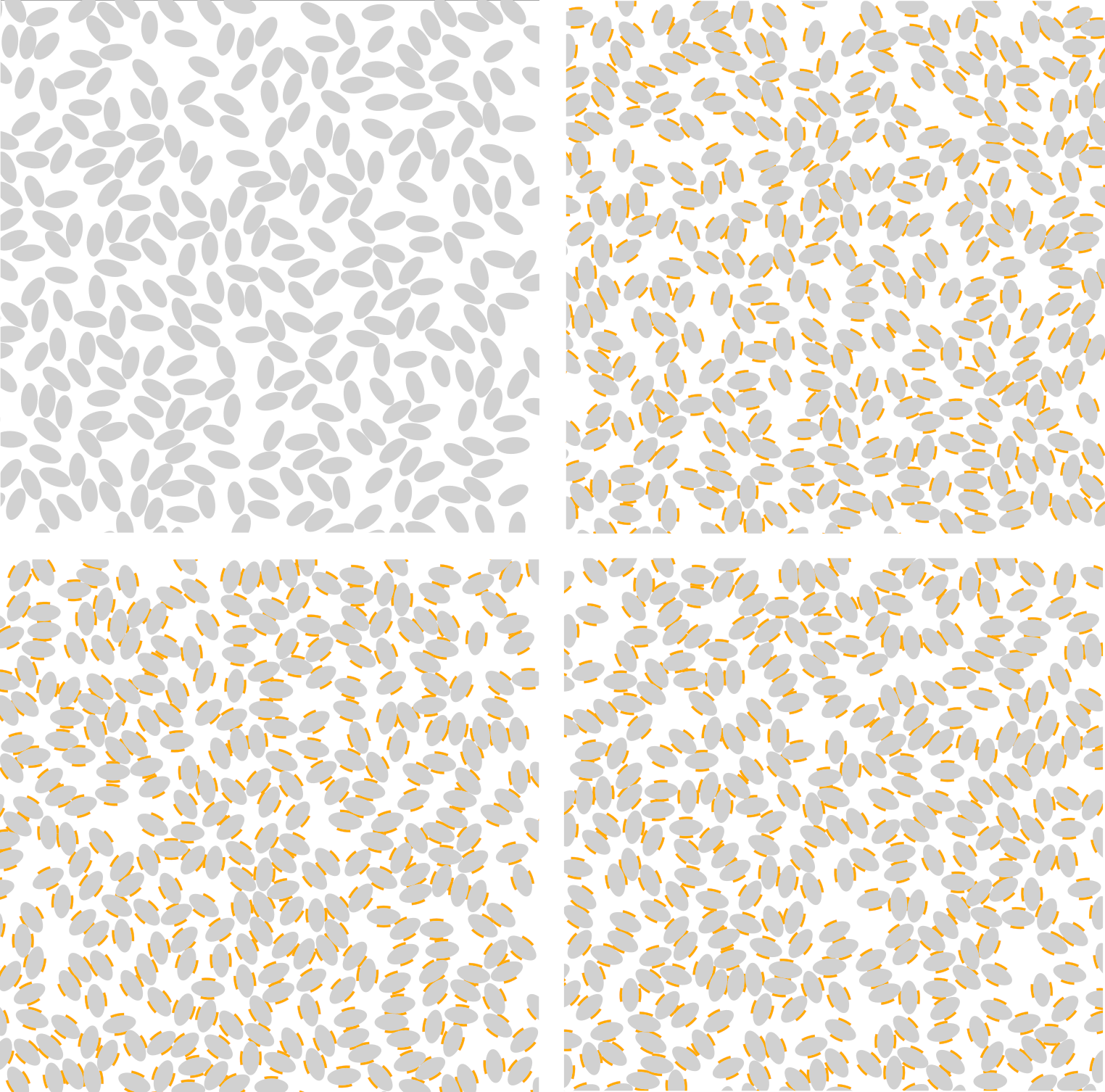}} 
\caption{Representative snapshots taken from our NPT MC simulations,
  showing particles in a section of the simulation cell. Results have
  been obtained for patchy elliptic particles with aspect ratio
  $\kappa = 2$ and reduced, dimensionless pressure $P^\star = \pi a b
  P/(k_{\rm B}T) = 2$. Selected temperatures are: $T^\star = \infty$
  -- top left, $T^\star = 2.0$ -- top right, $T^\star = 1.0$ -- bottom
  left, and $T^\star = 0.6$ -- bottom right. }
\label{fig:snap_2}
\end{figure}

\begin{figure}
\centering
\resizebox*{14cm}{!}{\includegraphics{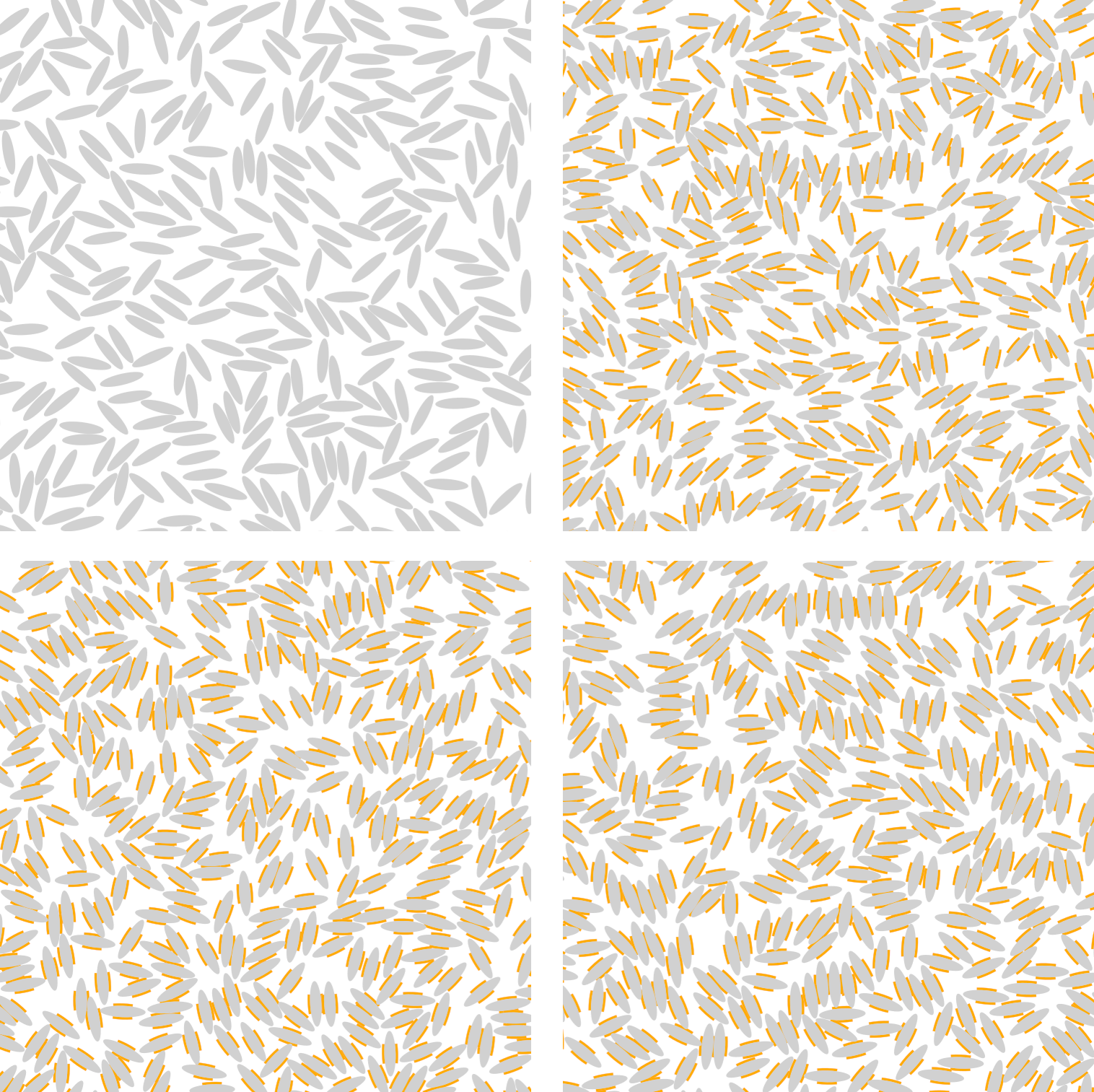}} \\
\caption{Representative snapshots taken from our NPT MC simulations,
  showing particles in a section of the simulation cell. Results have
  been obtained for patchy elliptic particles with aspect ratio
  $\kappa = 4$ and reduced, dimensionless pressure $P^\star = \pi a b
  P/(k_{\rm B}T) = 2$. Selected temperatures are: $T^\star = \infty$
  -- top left, $T^\star = 2.0$ -- top right, $T^\star = 1.0$ -- bottom
  left, and $T^\star = 0.6$ -- bottom right.}
\label{fig:snap_4}
\end{figure}

\begin{figure}
\centering
\resizebox*{14cm}{!}{\includegraphics{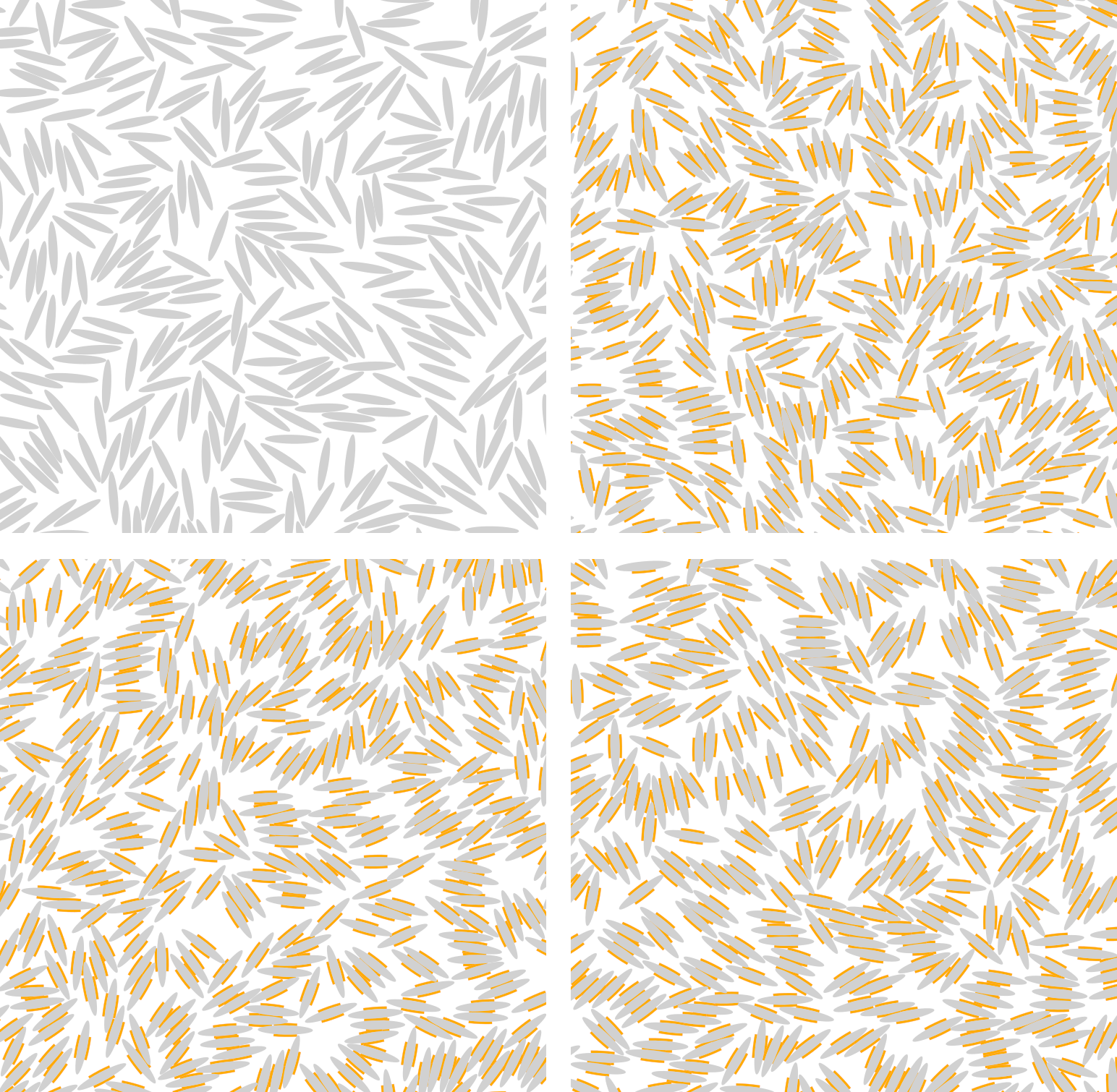}} 
\caption{Representative snapshots taken from our NPT MC simulations,
  showing particles in a section of the simulation cell. Results have
  been obtained for patchy elliptic particles with aspect ratio
  $\kappa = 6$ and reduced, dimensionless pressure $P^\star = \pi a b
  P/(k_{\rm B}T) = 2$. Selected temperatures are: $T^\star = \infty$
  -- top left, $T^\star = 2.0$ -- top right, $T^\star = 1.0$ -- bottom
  left, and $T^\star = 0.6$ -- bottom right.}
\label{fig:snap_6}
\end{figure}

We start our discussion with a qualitative inspection of selected
simulation snapshots, carried out for the three selected
$\kappa$-values and keeping the pressure $P^\star$ fixed.  Four
different temperatures were considered, including the case $T^\star =
\infty$; selected snapshots are shown in
Figs. \ref{fig:snap_2}, \ref{fig:snap_4}, and \ref{fig:snap_6}. As
expected, the attraction between the patches leads to the formation of
small aggregates as the temperature is decreased: the location of the
patches at the co-vertices of the ellipses induces a parallel
alignment of bonded particles. For $\kappa = 2$ we observe
even for the lowest temperature $T^\star = 0.6$ that only a few pairs of
particles form, while already for $\kappa = 4$ and, in particular, for
$\kappa = 6$ ``bundles'' of particles form.

\begin{figure}
\centering
\resizebox*{6.5cm}{!}{\includegraphics{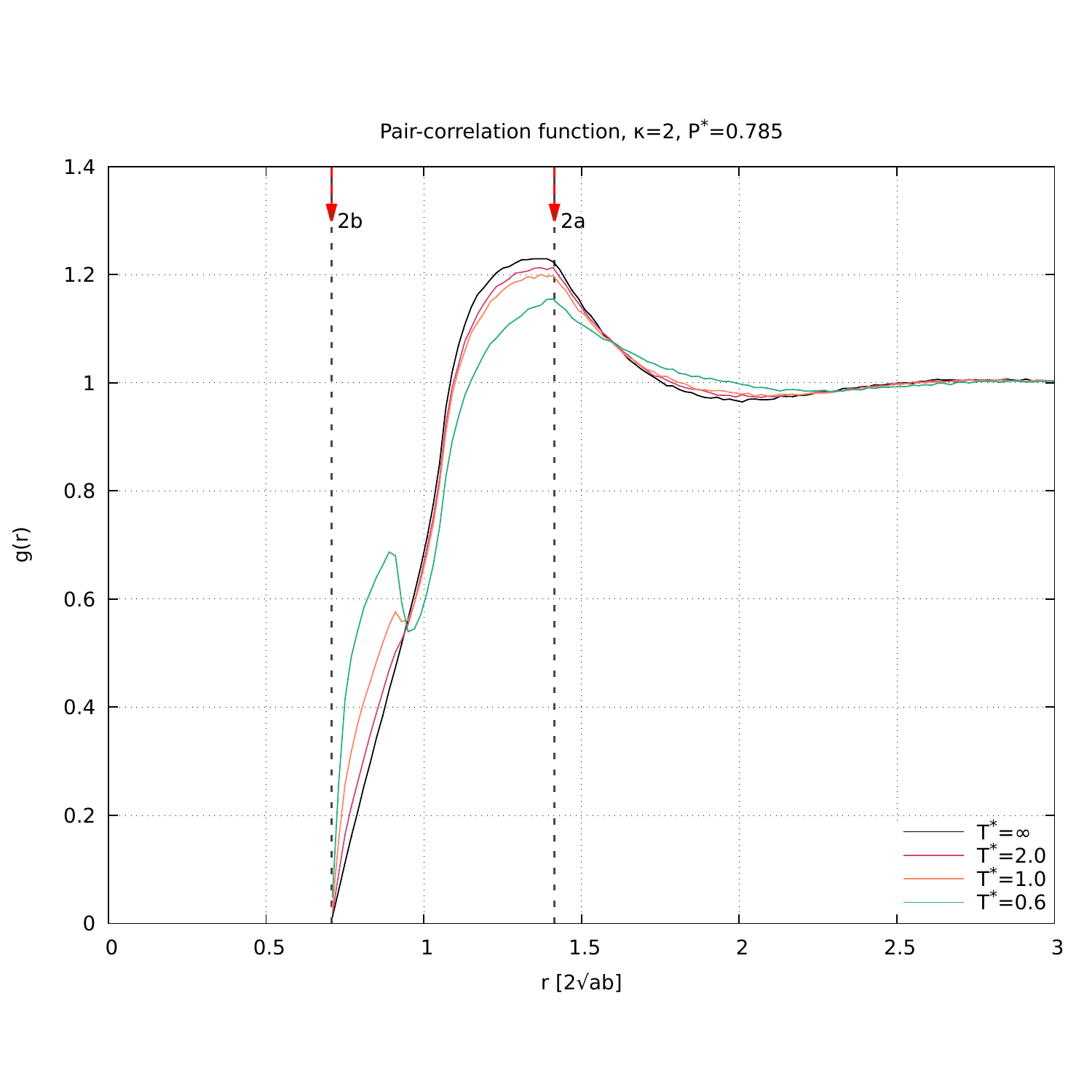}} 
\resizebox*{6.5cm}{!}{\includegraphics{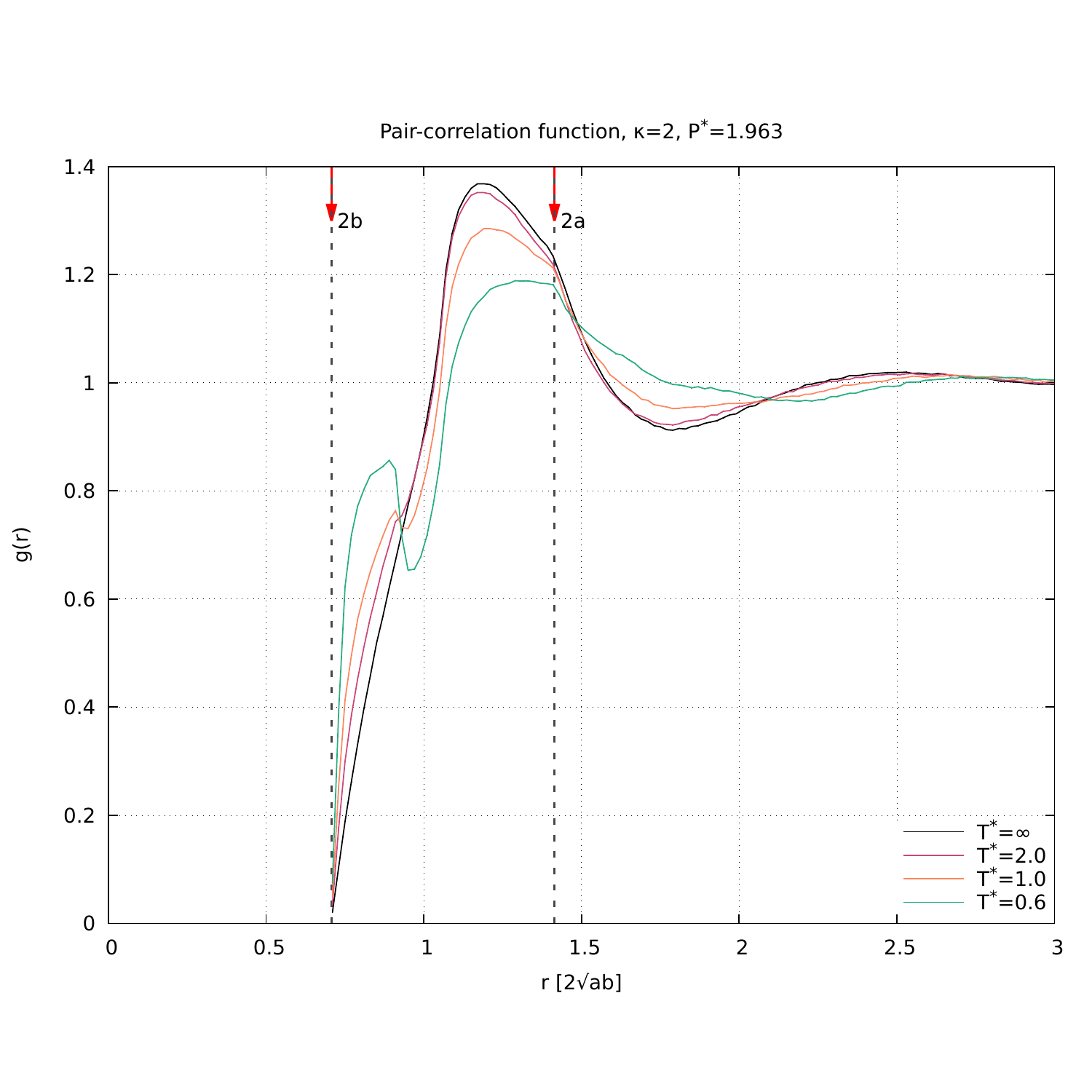}} \\
\resizebox*{6.5cm}{!}{\includegraphics{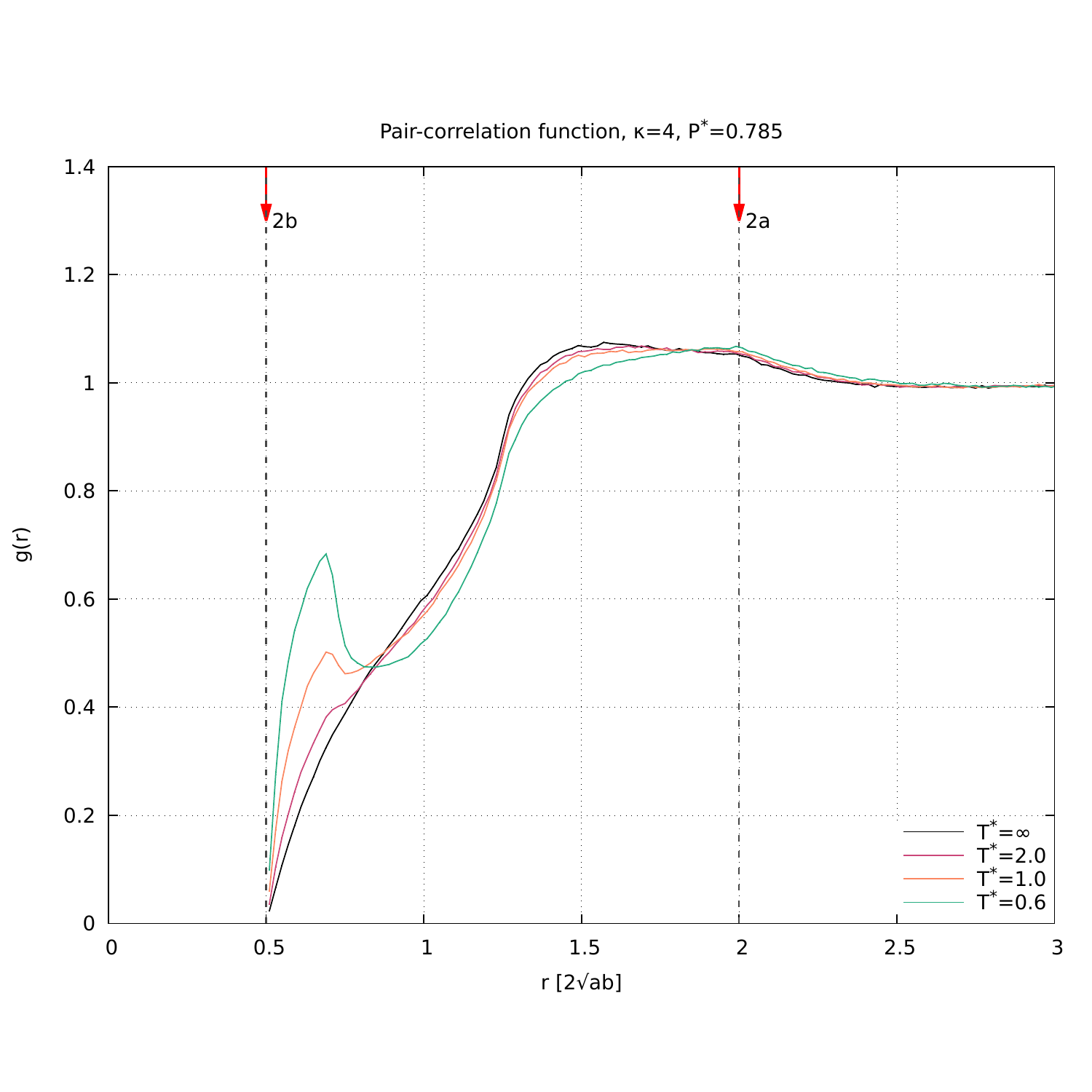}} 
\resizebox*{6.5cm}{!}{\includegraphics{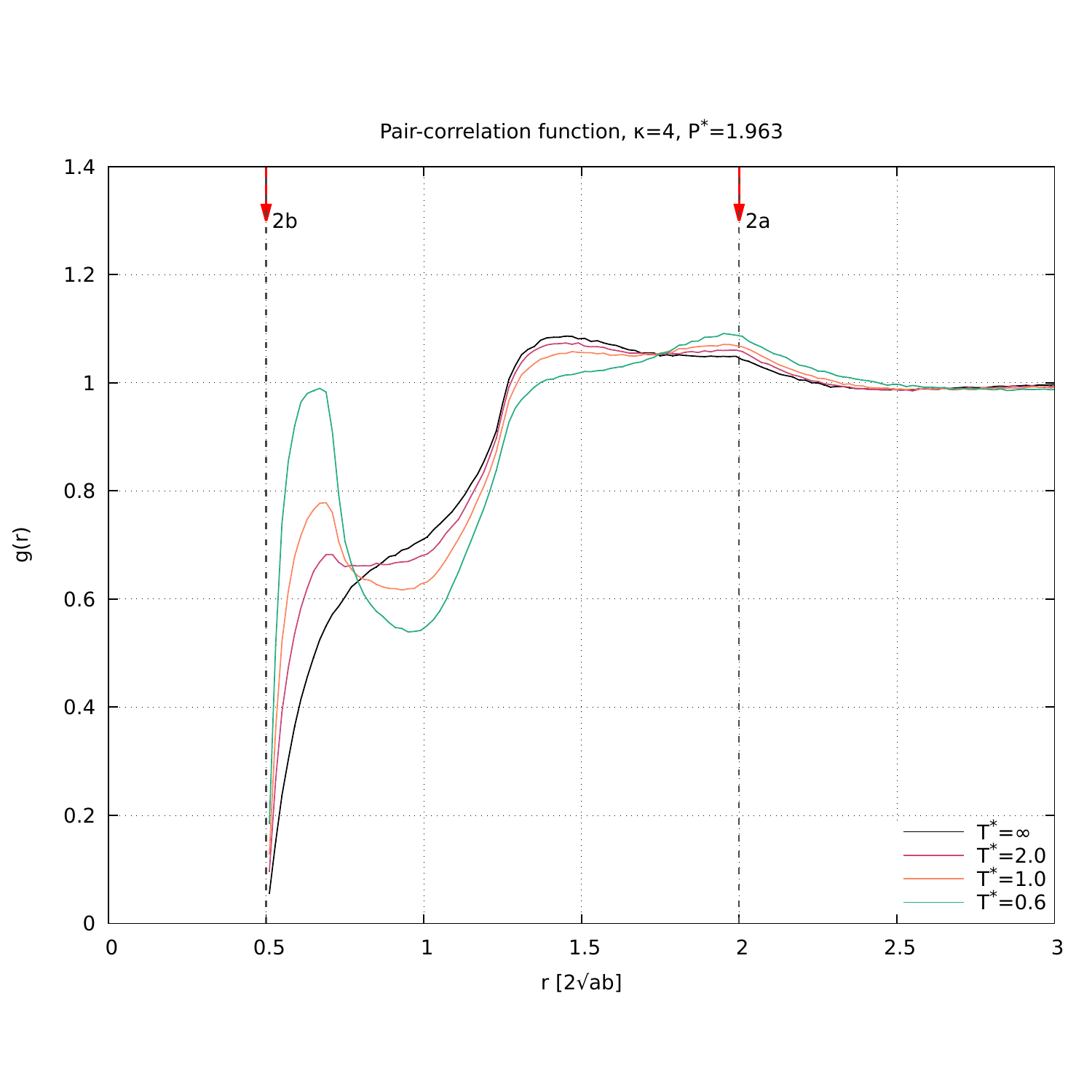}} \\
\resizebox*{6.5cm}{!}{\includegraphics{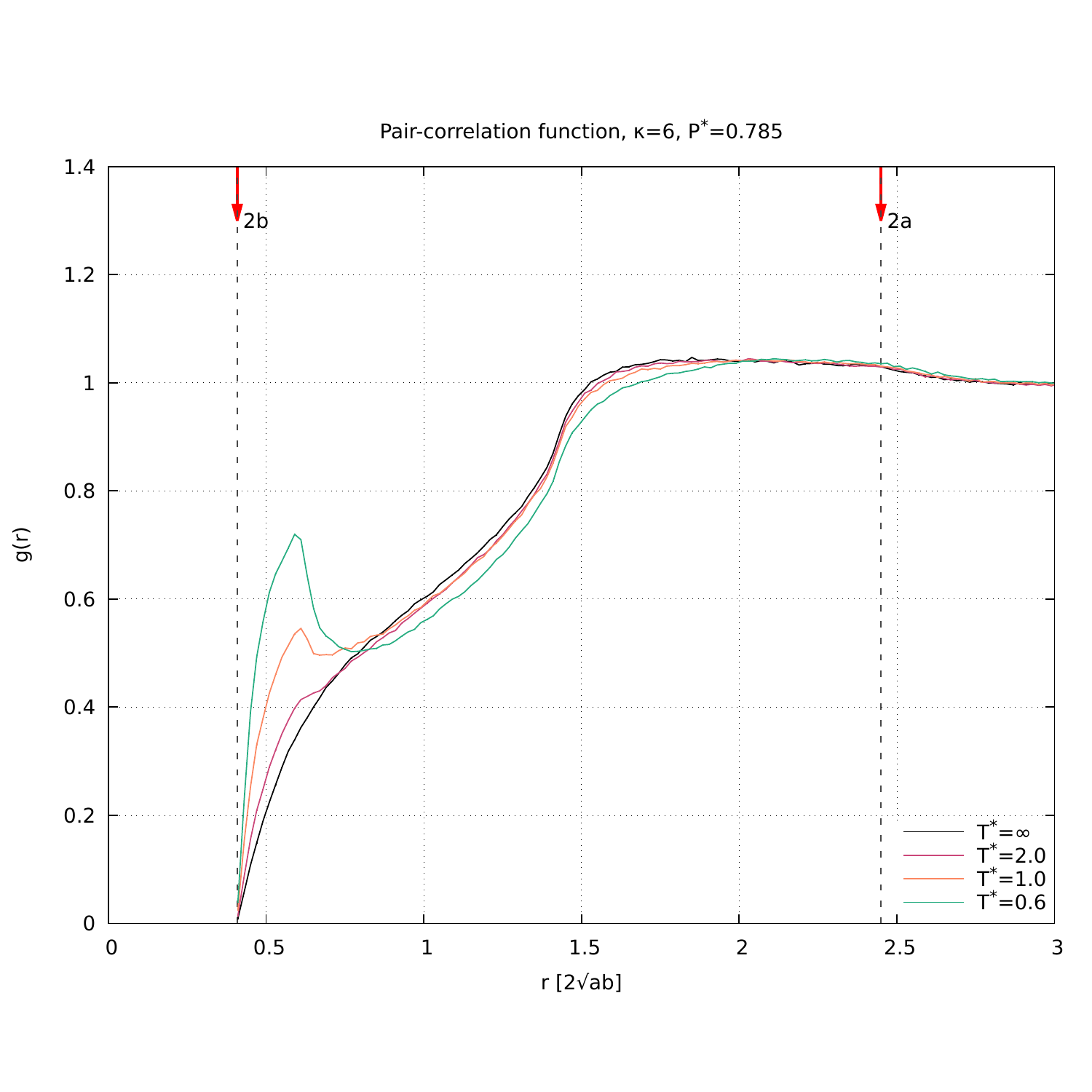}} 
\resizebox*{6.5cm}{!}{\includegraphics{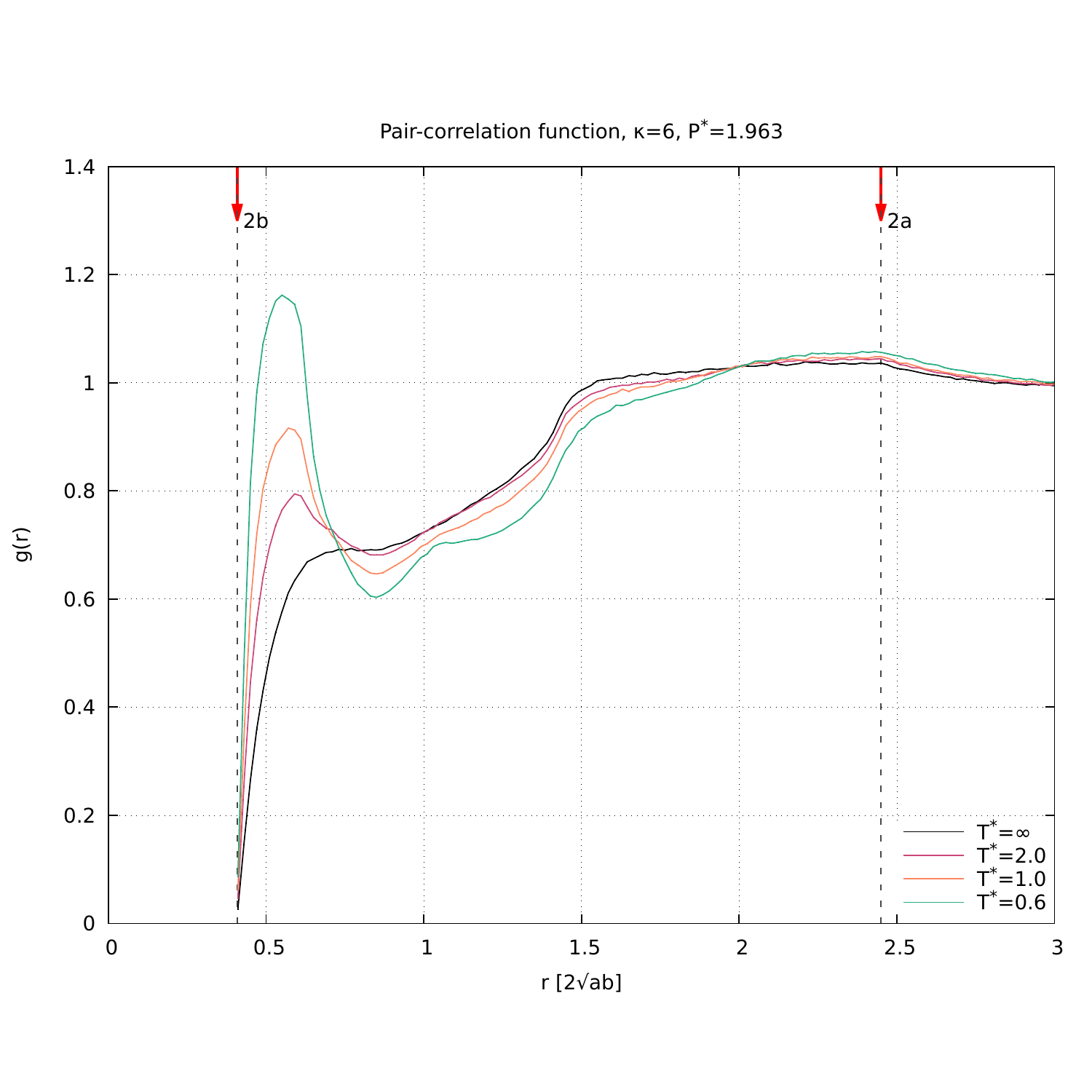}} \\
\caption{Radial distribution functions $g(r)$ as functions of $r$ (in
  reduced units of $2 \sqrt{ab}$) for $\kappa$- and $P^\star$-values
  as specified on top of each panel and for different temperatures (as
  labeled in the panels). The vertical arrows in each of the panels
  indicate for the respective $\kappa$-values the distances $r = 2a$
  and $r = 2b$ (again in reduced units $2 \sqrt{ab}$).}
\label{fig:pdf}
\end{figure}

We have analysed the spatial structure of the emerging
  disordered phases on a semi-quantitative level in terms of the
  conventional pair correlation function, $g(r)$, \cite{Hansen:2013};
  this function is depicted for the three selected values of $\kappa$
  and for representative values for $P^\star$ and $T$ in
  Fig. \ref{fig:pdf}. The results provide evidence that the attractive
  patch- patch interaction of the particles enhances the correlation
  between the particles (in particular as the temperature decreases
  and/or as the pressure increases); this trend is reflected by the
  emergence of a peak for distances close to $2b$, which grows with
  decreasing temperature at the cost of the main peak which is located
  at higher distances. The emergence of the first peak reflects the
  above mentioned bundle formation of the particles under these
  external conditions. Further, the shift of the position of the
  former main peak towards $r \simeq 2b$ indicates a stacking tendency
  of the bundles. A more quantitative analysis of the positional {\it
    and} of the {\it orientational} order of the particles requires
  the analysis of the data in terms of correlation functions
  $g_{2l}(r)$, as they have been defined in
  \cite{Cuesta:1990}. However, such an analysis has to be based on
  considerably more extensive simulations on a finer pressure- and
  temperature-grid and will be postponed to future work
  \cite{Wagner:2018}.

\subsection{Equation-of-state}
\label{subsec:eos}

\begin{figure}
\centering
\resizebox*{11.5cm}{!}{\includegraphics{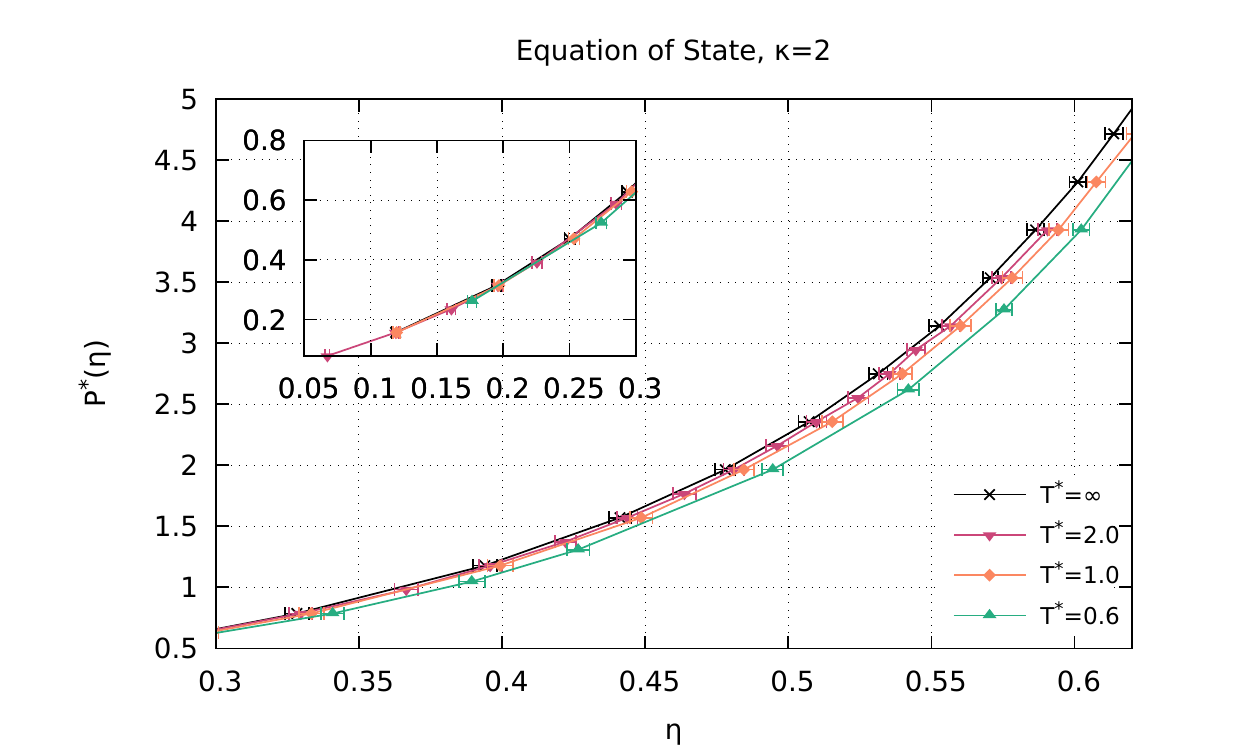}} 
\resizebox*{11.5cm}{!}{\includegraphics{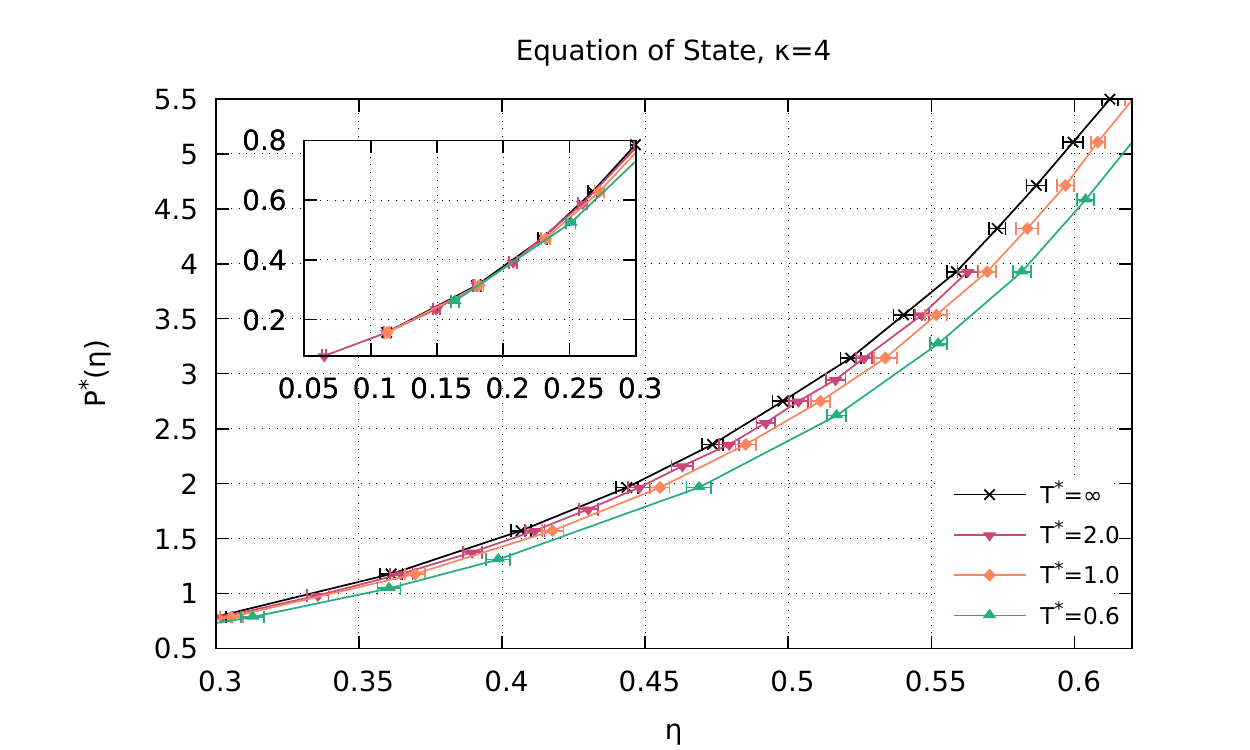}} \\
\resizebox*{11.5cm}{!}{\includegraphics{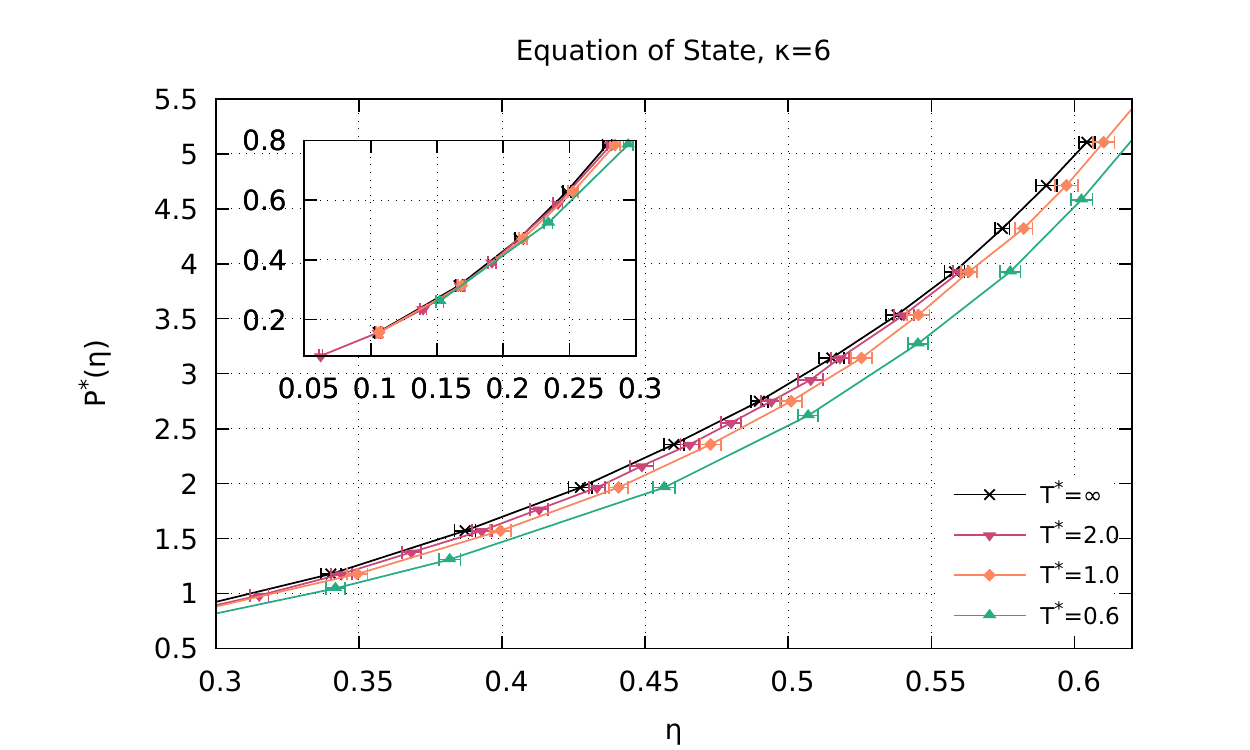}} 
\caption{Equation-of-state, i.e., $P^\star = \pi a b P/(k_{\rm B}T)$
  as a function of the packing fraction $\eta$ for four different
  temperatures (as labeled), including the case of undecorated, hard
  ellipses (i.e., for $T^\star = \infty$; in the latter case our results are
  consistent with the data presented in \cite{Cuesta:1990}. Results
  are shown for three different $\kappa$-values: top panel -- $\kappa
  = 2$, central panel -- $\kappa = 4$, and bottom panel -- $\kappa =
  6$. The insets show the trivial low-density equation-of-state of the
  system.}
\label{fig:EOS}
\end{figure}

In Fig. \ref{fig:EOS} we show the data for the equation-of-state,
i.e., $P^\star$ as a function of the packing fraction $\eta$, for the
three different $\kappa$-values investigated in this study. As
expected, the results for $P^\star$ tend with $T^\star \to \infty$ to the
data of the undecorated, hard ellipses. We observe a stronger decrease
in the pressure with decreasing temperature for larger
$\kappa$-values, reflecting the fact that for these system parameters
the particles show a stronger tendency to form clusters (see
discussion above). Data provide -- in particular for $\kappa = 6$ --
evidence of fingerprints of an emerging nematic phase in the
equation-of-state. Investigations dedicated to the ordered phases of
the system are envisaged; they require, however, more extensive
simulations and more sophisticated simulation techniques as, for
instance, NPT simulations with a variable box shape. Preliminary
results indicate that such simulations are very time consuming, in
particular for larger $\kappa$-values, (inducing strong deformations
of the simulation box).

\section{Conclusion}
\label{sec:conclusion}

In this contribution we have introduced a simple model for an {\it
  anisotropic} patchy particle. Its elliptic, impenetrable body (with
aspect ratio $\kappa$) is decorated on its (co-)vertices by up to four
Kern-Frenkel type of patches, which are shaped via two concentric
circles, the inner ones being the osculating circles of the respective
(co-)vertex; the radius of the outer circle sets the range of
interaction which is assumed to be constant within the patch (i.e., a
square-well type of interaction).  Using (i) well-known criteria that
provide evidence whether two ellipses overlap, touch each other, or do
not overlap and (ii) the conventional criterion put forward within the
original Kern-Frenkel model for the patch-patch interaction of
circular particles our model is easily amenable to MC simulations.

We have studied elliptic patchy particles of aspect ratios $\kappa =
2, 4$, and 6, decorating the particles on their co-vertices with
patches. Investigations have been carried out via standard Monte Carlo
simulations in the NPT ensemble, considering typically $\sim$~1000
particles.

Even though we have to postpone a more detailed and quantitative
analysis of the structural data of the emerging phases to a future
publication, we observe a clear tendency that the degree of shape
anisotropy (in terms of the aspect ratio $\kappa$) has a strong
influence on the phase formation of the particles and on their
equation-of-state: (i) comparable external parameters (i.e.,
temperature and pressure) lead to pronounced ordering effects (i.e.,
bundles of aligned, interacting particles) as $\kappa$ increases,
while for $\kappa = 2$ the formation of this type of order sets in
only at considerably higher pressure-values; (ii) the patch-patch
attraction leads to the expected shift in the equation-of-state (i.e.,
of the pressure) to higher densities. A more detailed analysis of the
emerging phases and the location of the related phase boundaries
requires more extensive simulations on a considerably finer pressure-
and temperature-grid; we postpone these investigations to a future
contribution.

Despite its simplicity, the model captures both shape anisotropy and
patchiness in a faithful manner. It is also versatile in the sense as
it is easily amenable to MC simulations and -- using a continuous
patch interaction instead of the square-well type of potential -- is
also ready to be used in standard molecular dynamics simulations. Via
the different variants of surface decoration it is also possible to
feature chirality by considering left- and right-handed particles in a
binary mixture. The impact of the competition between attraction and
repulsion or the competition of patches of different spatial extent on
the emerging phases are other features that we plan to investigate in
the future.

\section*{Acknowledgement(s)}
Financial support by the TU Wien Doctorate College ``BIOINTERFACE'' is
gratefully acknowledged. Computational resources by the Vienna
Scientific Cluster (VSC) are gratefully acknowledged.

\end{document}